\pdfoutput=1 
\documentclass{aa}  

\usepackage{amssymb}
\usepackage{euscript}
\usepackage{balance}
\usepackage{amsmath}
\usepackage{graphicx}
\usepackage{txfonts}
\usepackage{multirow}
\begin{document} 

   \title{Spot cycle reconstruction: an empirical tool}
   \subtitle{Application to the sunspot cycle}

   \author{A. R. G. Santos \inst{1,2,3,4}
          \and M. S. Cunha \inst{1,2,3}
          \and P. P. Avelino \inst{1,2,3}
           \and T. L. Campante \inst{4}}

   \institute{Instituto de Astrof\'{i}sica e Ci\^{e}ncias do Espa\c{c}o, Universidade do Porto, CAUP, Rua das
Estrelas, 4150-762 Porto, Portugal\\
              \email{asantos@astro.up.pt}
              \and Centro de Astrof\'{i}sica da Universidade do Porto, Rua das Estrelas, 4150-762 Porto, Portugal
              \and Departamento de F\'{i}sica e Astronomia, Faculdade de Ci\^{e}ncias, Universidade do Porto,
Rua do Campo Alegre 687, 4169-007 Porto, Portugal
             \and School of Physics and Astronomy, University of Birmingham, Edgbaston, Birmingham B15 2TT, UK}

  \date{Received November 7, 2014 / Accepted June 4, 2015}
 
  \abstract
   {The increasing interest in understanding stellar magnetic activity cycles is a strong motivation for the development of parameterized starspot models which can be constrained observationally.}
   {In this work we develop an empirical tool for the stochastic reconstruction of sunspot cycles, using the average solar properties as a reference.}
   {The synthetic sunspot cycle is compared with the sunspot data extracted from the National Geophysical Data Center, in particular using the Kolmogorov-Smirnov test. This tool yields synthetic spot group records, including date, area, latitude, longitude, rotation rate of the solar surface at the group's latitude, and an identification number.}
   {Comparison of the stochastic reconstructions with the daily sunspot records confirms that our empirical model is able to successfully reproduce the main properties of the solar sunspot cycle. As a by-product of this work, we show that the Gnevyshev-Waldmeier rule, which describes the spots’ area-lifetime relation, is not adequate for small groups and we propose an effective correction to that relation which leads to a closer agreement between the synthetic sunspot cycle and the observations.}
   {}


  \maketitle
%

\section{Introduction}

In the Sun, the level of magnetic activity varies over time showing a periodic behaviour known as the solar cycle. The most direct evidence for this is the 11-year variation in the number of sunspots observed at the solar photosphere. As the cycle proceeds, the sunspot formation zone gradually migrates towards the equator until the next minimum is reached and a new cycle begins. At that point, the polarity of the magnetic field reverses and returns to the original state at the end of the second 11-year cycle, thus completing a 22-year Hale cycle \citep[the sunspot polarity law -][]{hale19,hale25}.

Evidence for the presence of activity cycles in other solar-like stars, including exoplanet hosts, has also been accumulating over the past years. Starspots cannot be observed directly at the surface of distant stars, but can be detected indirectly through the effects they induce, such as the strong emission at the centre of the \ion{Ca}{II} H and K lines. By monitoring this emission, it is possible to infer the rotation period of the star and the period of its magnetic cycle \citep[e.g.][]{wilson,duncan,gray,saar,cincunegui,hall,metcalfe10,metcalfe13}. Moreover, since spots are darker than the average stellar surface they can also be detected through the inspection of the photometric light curves of stars observed by CoRoT \citep[][]{baglin} and Kepler \citep[][]{borucki,kock} space missions. In an active star with starspots crossing the visible disk, the light curve shows a quasi-periodic modulation which results from the combination of the effects of stellar rotation and magnetic activity \citep[e.g.][]{mosser09,mathur10,garcia10ax,ballot,campante11}. That modulation is more significant in periods of maximum activity, making the light curves a possible starspot proxy. In addition, through the wavelet analysis of the light curves it might be possible to obtain information about the time evolution of the starspots at the stellar surface and, in turn, derive the period of the activity cycle and the rotation period of the star \citep[e.g.][]{campante12,mathur14,garcia14,bravo}.Finally, stellar activity cycles may also be detected through their impact on stellar oscillations because the magnetic activity affects the wave propagation, inducing changes to the oscillation frequencies, amplitudes, and line widths, which are thus found to vary in phase with other sun-/starspot proxies \citep[e.g.][]{woodard,libbrecht,chaplin04,metcalfe07,garcia10,tripathy11}.

Starspots simulations can be used to study how the activity-induced variability in the light curve and in the oscillation properties depends on the characteristics of the activity cycle. These simulations are important in other astrophysical contexts, such as in the quest for exoplanets, where they can be used to design new strategies to reduce the signatures induced by stellar activity on the radial velocity and transit observations \citep[e.g.][]{pont,czesla,figueira,dumusque,oshagh13,oshagh14}. With the above in mind, in this work we develop a parameterized model aimed at reproducing the main properties of the sunspot cycle that can also be applied to reproducing the activity cycles of other stars. In order to build our empirical model, we use as inputs a number of properties of the observed sunspot cycle. These properties are summarized in section \ref{sec:properties} and their implementation in our model is discussed in section \ref{sec:model}. The results obtained with this tool and their comparison with the solar data extracted from the National Geophysical Data Center (NOAA/NGDC) are presented in section \ref{sec:results}. Finally, in section \ref{sec:conclusions} we draw our main conclusions.

\section{Properties of the solar cycle}\label{sec:properties}

In order to reproduce the sunspot cycle, a number of important observational constraints must be considered. In what follows we review key observational properties of the sunspot cycle that will be used in our model, namely: the number of sunspot groups and its dependence on the phase of the cycle; the sunspots' areas and their relation with the sunspots' lifetimes; the formation latitude of the sunspots and the width of the formation region; the differential rotation of the solar surface.

The number of sunspots and sunspot groups varies over the solar cycle. Its evolution is asymmetric: the rising phase of the cycle is faster than the declining towards the next minimum. Different authors have used different functions to describe the asymmetric shape of the sunspot cycle \citep[e.g.][]{setwart,elling,sabarinath,volobuev,du}. In particular, \citet*{hathaway94} found that the observed number of sunspots is nicely fitted by the following function of time,
\begin{equation}
N_\textrm{S}(t)=\dfrac{a_1(t-t_0)^3}{\exp((t-t_0)^2/b_1^2)-c_1}\,,\label{eq:f1}
\end{equation}
where $t_0$ is the starting time (about four months prior to the minimum for an average cycle \citep[][]{hathaway10}), $a_1$ is the amplitude, $b_1$ is related to the size of the cycle, and $c_1$ is related to the asymmetry of the cycle. 

The asymmetric shape of the solar cycle is also evident in the temporal variation of the sunspot areas.
\citet{bogdan} were the first to notice that the accumulated umbral areas distribution (derived from daily records by counting each spot as many times as the number of days it remains visible) can be described by a log-normal distribution. Since the ratio between the umbral area and the total area of the spot does not depend significantly on the spot size \citep[e.g.][]{brandt,solanki03,vaquero,kiess}, the accumulated spot areas also follow a log-normal distribution. This finding was confirmed by later studies \citep[e.g.][]{baumann,hathaway08,kiess}. Moreover, \citet{baumann} have shown that the log-normal distribution also nicely fits the observed distributions for the instantaneous area and for the maximum area of sunspots. However, according to \citet{jiang}, the area distribution for groups smaller than 60 MSH (millionth of the solar hemisphere) is better described by a power law.

During its life, a given sunspot group grows until it reaches a maximum area and then decays. The growth ($\Psi$) and decay ($\Gamma$) rates, i.e. the time derivative of the group's area during each of these phases, are found to be dependent on the group areas, the activity cycle, the phase of the cycle, and the latitude \citep[e.g.][]{moreno,howard92,petrovay,hathaway08,javariah}. Small groups grow faster than they decay, while the growth rates of large groups are smaller than their decay rates \citep[][]{howard92}. \citet{hathaway08} found a linear relation between the decay rates and the group area, but the erosion model proposed earlier by \citet{petrovay} indicates a non-linear relation of the type $\Gamma\propto A^{0.5}$, where $A$ is the group area at a given time. More recently, \citet{javariah} suggested that the relation between the decay rates and areas could be better described by a power law of the form $\Gamma=\exp(\gamma_1)A^{\gamma_2}$. The constant $\gamma_2$ was found to vary from $\sim0.45$ to $\sim0.70$, when considering individual cycles and different phases of each cycle (with $A$ expressed in MSH). When assuming the sunspot data from 1874-2011, \citet{javariah} found that $\gamma_1\sim0.26$ and $\gamma_2\sim0.613$. On the other hand, the same study did not produce conclusive results regarding the relation between the group's growth rate and its area.

The areas of sunspots and sunspot groups are also related to their lifetimes, which can range from hours to months, depending on their size. This dependency is described by the Gnevyshev-Waldmeier (GW) rule \citep{gnevyshev,waldmeier}, according to which,
\begin{equation}
A_m=D_\textrm{GW}T\,.
\end{equation}
Here, $A_m$ is the sunspot or sunspot group maximum area (in MSH),  $T$ is the corresponding lifetime, and $D_\textrm{GW}$ is a constant of proportionality (around 10\,MSH\,day$^{-1}$). The determination of a precise value for $D_\textrm{GW}$ is hampered by the difficulty in measuring the spots' lifetimes due to the nightfall, the solar rotation (lack of observations of the invisible side of the Sun), and limb darkening \citep[e.g.][]{henwood,blanter,solanki03}. In spite of these difficulties, some studies have been carried out, indicating that $D_\textrm{GW}$ might be larger than first estimated. \citet{petrovay} found that $D_\textrm{GW}$ is $10.89\pm0.18$\,MSH\,day$^{-1}$ for individual sunspots. More recently, \citet{henwood} studied long-lived sunspot groups and estimated that $D_\textrm{GW}=11.73\pm0.26$\,MSH\,day$^{-1}$.

As was first reported by \citet{carrington}, the sunspot formation latitude also varies periodically with time. At the beginning of a new cycle the first spots appear at latitudes of about $\pm 40^{\circ}$. The succeeding spots form at progressively lower latitudes, being rarely observed within $\pm 5^{\circ}$. At the solar minimum the last spots of the cycle emerge at low latitudes, while spots of the new cycle start to form at high latitudes. This behaviour is known as the Sp\"{o}rer law and it may be seen in the butterfly diagram \citep[or Maunder diagram;][]{maunder}.
Despite the presence of short plateaus at intermediate latitudes \citep[$\sim 10^{\circ}$ - around maximum;][]{chang}, \citet{hathaway11} found that the drift of the sunspot zones follows an exponential function, where the average latitude, $L_\textrm{S}$, is given by 
\begin{equation}
L_\textrm{S}(t)=L_0\exp\left(-\dfrac{t-t_0}{7.5}\right)\,,\label{eq:hathaway11}
\end{equation}
where $L_0$ is the mean latitude at the time $t_0$, and $t$ is expressed in years. By considering the intermediate phases of the cycle where there is no overlap between consecutive cycles, \citet{jiang} verified that the evolution of the average latitude can also be described by a second-order polynomial.

\citet{chang} found that the spatial distribution of the sunspot groups at each time $t$ is bimodal and that it can be described by a double Gaussian; instead, \citet{ivanovb} showed that one single Gaussian describes the data reasonably well.
Moreover, the width of the sunspot formation zone, $\sigma_\textrm{L}$, also varies over the solar cycle \citep{gleissberg}. According to \citet*{miletskii09}, \citet{ivanovb}, and \citet{ivanova}, this width is a linear function of the activity level. However, according to \citet{jiang}, it can be described in relation to the average sunspot group latitude by a second-order polynomial of the form
\begin{equation}
\dfrac{\sigma_\textrm{L}}{L_\textrm{S}}=a_\sigma+b_\sigma \dfrac{t-t_{min}}{P_c}+c_\sigma \left(\dfrac{t-t_{min}}{P_c}\right)^2,
\end{equation}
where $a_\sigma$, $b_\sigma$, and $c_\sigma$ are the coefficients of the polynomial, $t_{min}$ corresponds to the minimum, and $P_c$ is the period of the cycle.

We note that sunspots are depressed (Wilson depression), thus they move according to the subphotospheric layers, i.e. slightly faster than the solar surface \citep[e.g.][]{zappala,zuccarello,abuzeid,schou,kitchatinov}. Since the spots' depths decrease as they evolve, younger spots also move faster than the older ones.

Finally, we note that the properties described above vary from cycle to cycle and some are found to be correlated \citep[e.g.][]{solanki02,solanki08,hathaway10,jiang}.
  
\section{Empirical solar cycle model}\label{sec:model}

The primary goal of this work is to produce a tool capable of reproducing an activity cycle that retains the main observational properties of the solar cycle. To that end, we develop an empirical model to generate sunspot groups as a function of time and gradually adapt the model assumptions and inputs until our goal is reached. To decide whether or not a given assumption/input is a better representation of the observational data than the previous ones we use the Kolmogorov-Smirnov test, which is described in section \ref{sec:KStest}. 

\subsection{Number of sunspot groups}\label{sec:nsp}

In our model, each sunspot group is generated independently from the others. This could not be assumed if we were considering individual spots, since the formation of spots within the same group is not independent. 

\begin{figure}[h!]
\centering
\includegraphics[width=9cm]{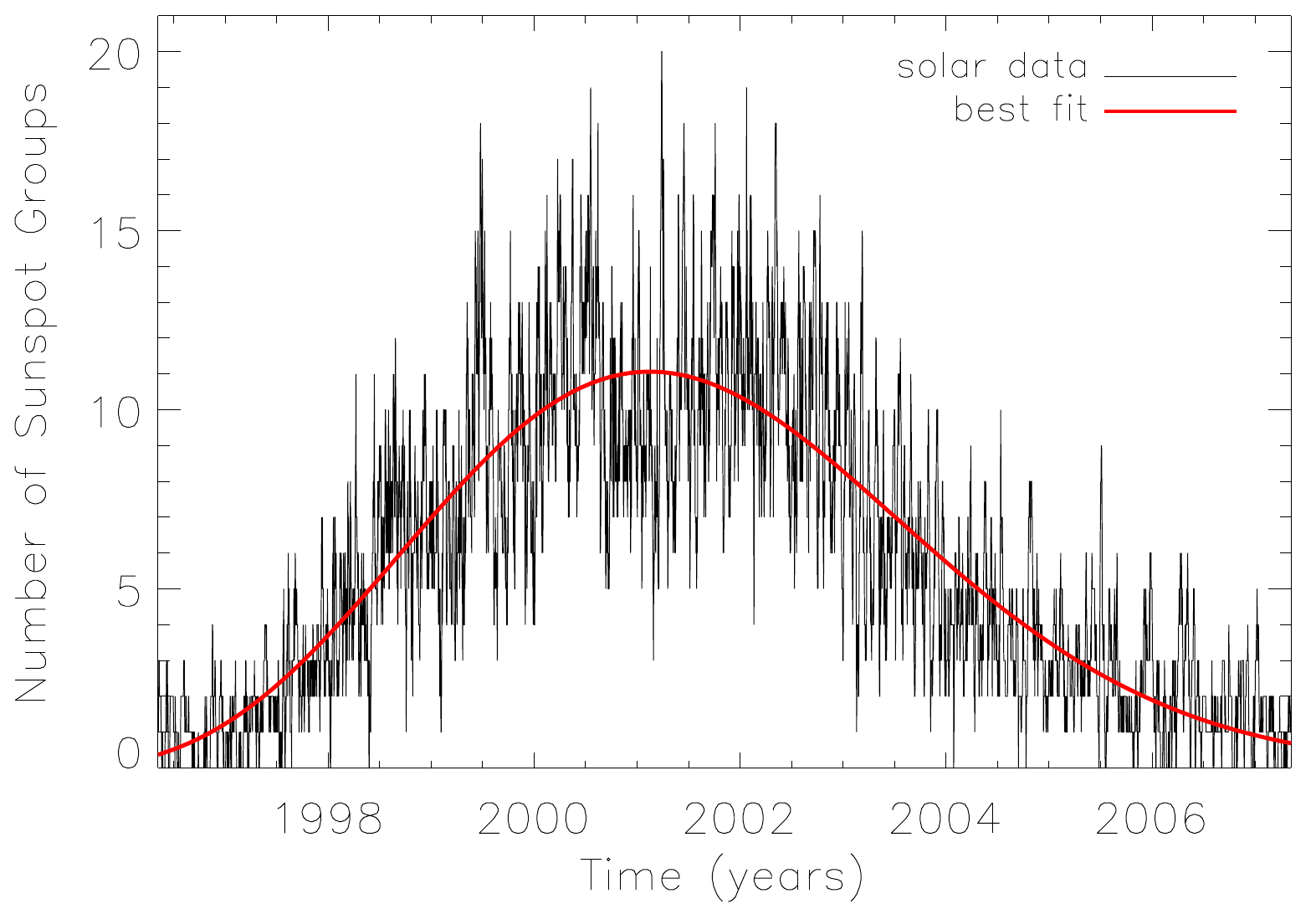}\caption{Variation of the number of sunspot groups over the solar cycle 23 (black). The red smooth line corresponds to the fit obtained using the function defined by \citet{hathaway94}. The observational data were extracted from NOAA.}\label{fig:hathaway}
\end{figure}

At each time step (fixed on one day to be comparable to the daily records of the sunspot data), $N$ groups are formed. The number of generated groups is randomly determined using a Poisson distribution with a mean value $N_\textrm{m}$ that depends on time.  In the current version of the model, $N_\textrm{m}$ is taken to be one sixth of $N_\textrm{s}$, where the function of time $N_\textrm{s}$ is derived from a fit of equation~(\ref{eq:f1}) to  the number of observed sunspot groups for solar cycle 23 (Fig. \ref{fig:hathaway}). With this choice for $N_\textrm{m}$, we find that the function $N_\textrm{s}$ derived from the fit to the synthetic data is always in reasonable agreement with that derived from the solar data.

Other ways to determine $N_\textrm{s}$ from the solar data were explored. The functions used by \citet{du} and \citet{sabarinath} (the latter takes into account the double peak feature of the solar cycle) led to results that are comparable to those obtained with equation (\ref{eq:f1}).

\subsection{Sunspot formation zone}\label{sec:lspots}

\begin{figure}[h!]
\centering
\includegraphics[width=8cm]{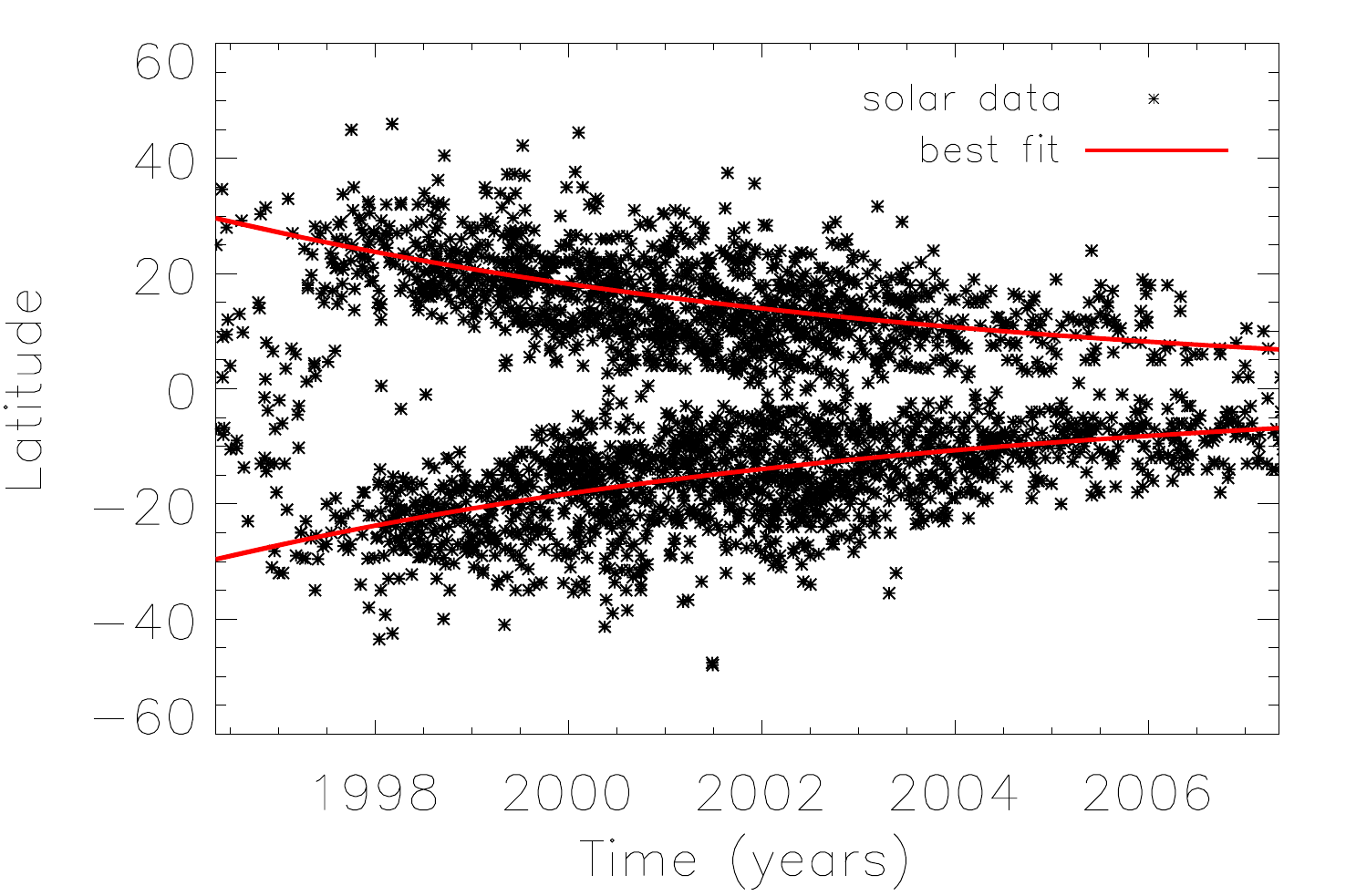}
\caption{Comparison between the latitudinal distribution of sunspots and the exponential fit of \citet{hathaway11}. The observational data was extracted from NOAA.}\label{fig:lspots}
\end{figure}

\begin{figure}
\centering
\includegraphics[width=7.3cm]{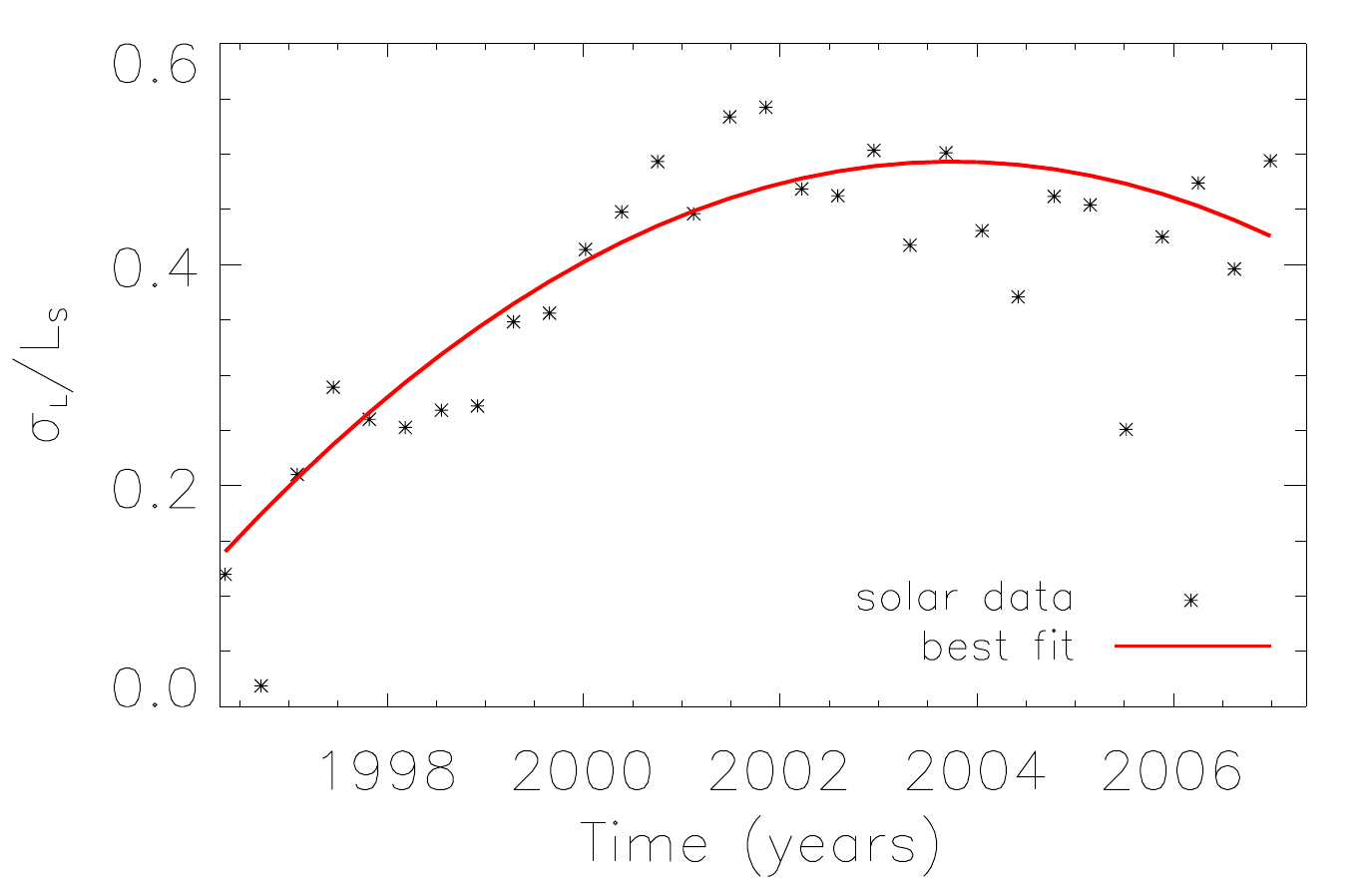}
\caption{Ratio between the standard deviation of the latitudinal distribution and the corresponding mean latitude (observational data from NOAA). The red solid line corresponds to the best fit.}\label{fig:sig}
\end{figure}

In our model, the latitude of each spot, $L$, is determined randomly through a single-Gaussian distribution, with a mean latitude $L_\textrm{S}$ and dispersion $\sigma_\textrm{L}$, both dependent on time. For $L_\textrm{S}$, we use Hathaway's exponential function given by equation (\ref{eq:hathaway11}) fitted to the northern hemisphere solar data for cycle 23 (Fig.~\ref{fig:lspots}), with $t_0$ fixed at the value found in section \ref{sec:nsp}. We opted to use solar data from a single hemisphere because in our model we do not account for the north-south asymmetry. For $\sigma_\textrm{L}$, we assume a second-order polynomial (Fig. \ref{fig:sig}), as suggested by \citet{jiang}. Other functions were considered to describe the latitudinal distribution of the sunspot groups (both for $L_\textrm{S}$ and $\sigma_\textrm{L}$). However, those led to a lower level of agreement between the synthetic and the observational data.

\subsection{Sunspot group areas and lifetimes}\label{sec:arealife}

\begin{figure}[h!]
\centering
\includegraphics[width=8.cm]{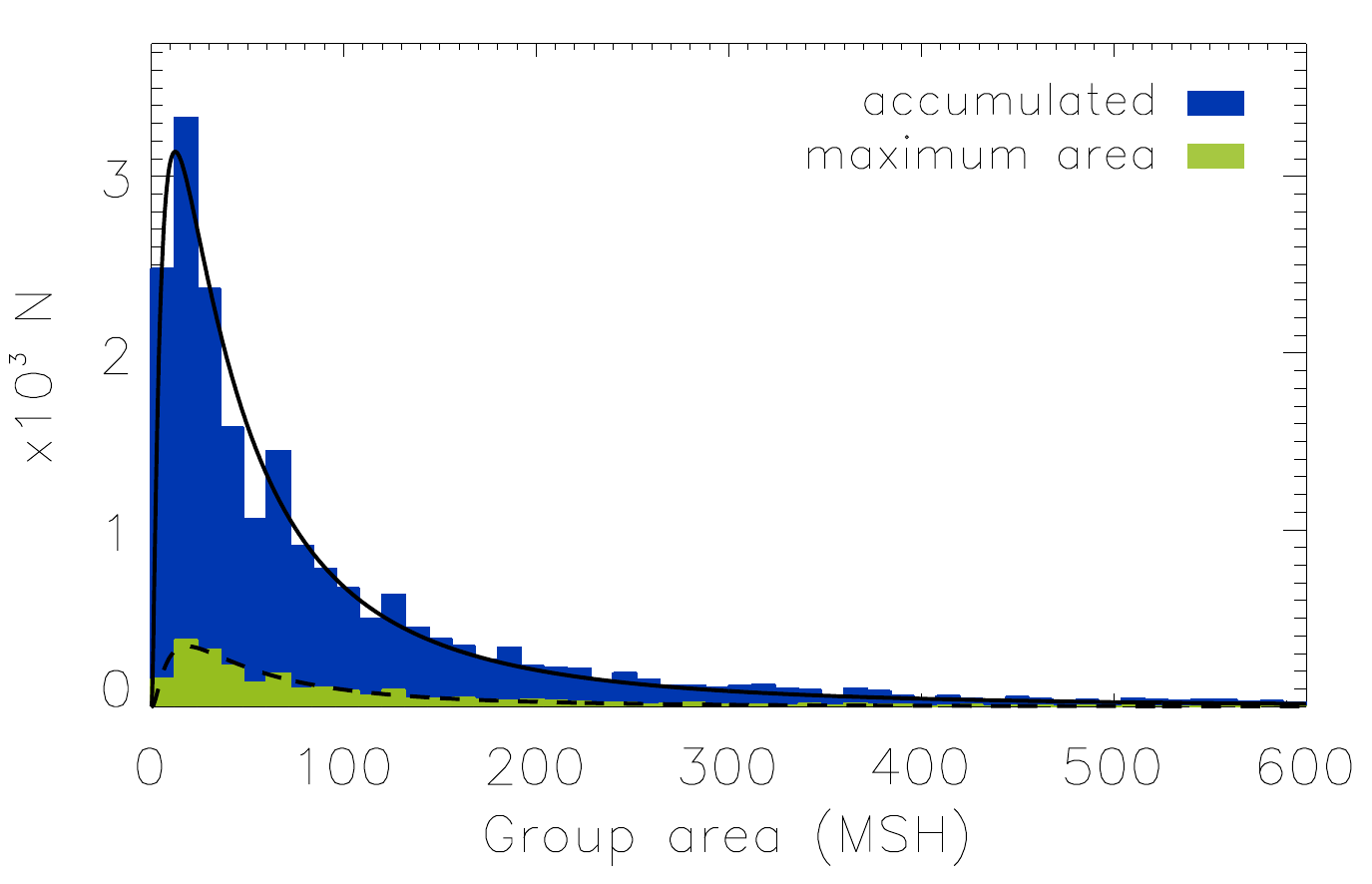}
\caption{Accumulated (blue) and maximum (green) area distributions for the solar data from NOAA. The black lines are the respective log-normal fits.}\label{fig:histtwo}
\end{figure}

Having defined the procedure to generate the number of sunspot groups and their position in latitude at each time step, we need to associate a maximum area ($A_\textrm{m}$) to each of them. Although the mean group area varies over the solar cycle \citep[e.g.][]{jiang}, we fix the area distribution. In accordance with the discussion in section \ref{sec:properties}, in our model the sunspot group maximum areas are drawn from a log-normal distribution whose parameters are obtained from a log-normal fit to the sunspot groups observations for cycle 23, considering each group only once and its maximum area (Fig.~~\ref{fig:histtwo}).

\begin{figure}
\centering
\includegraphics[width=8.cm]{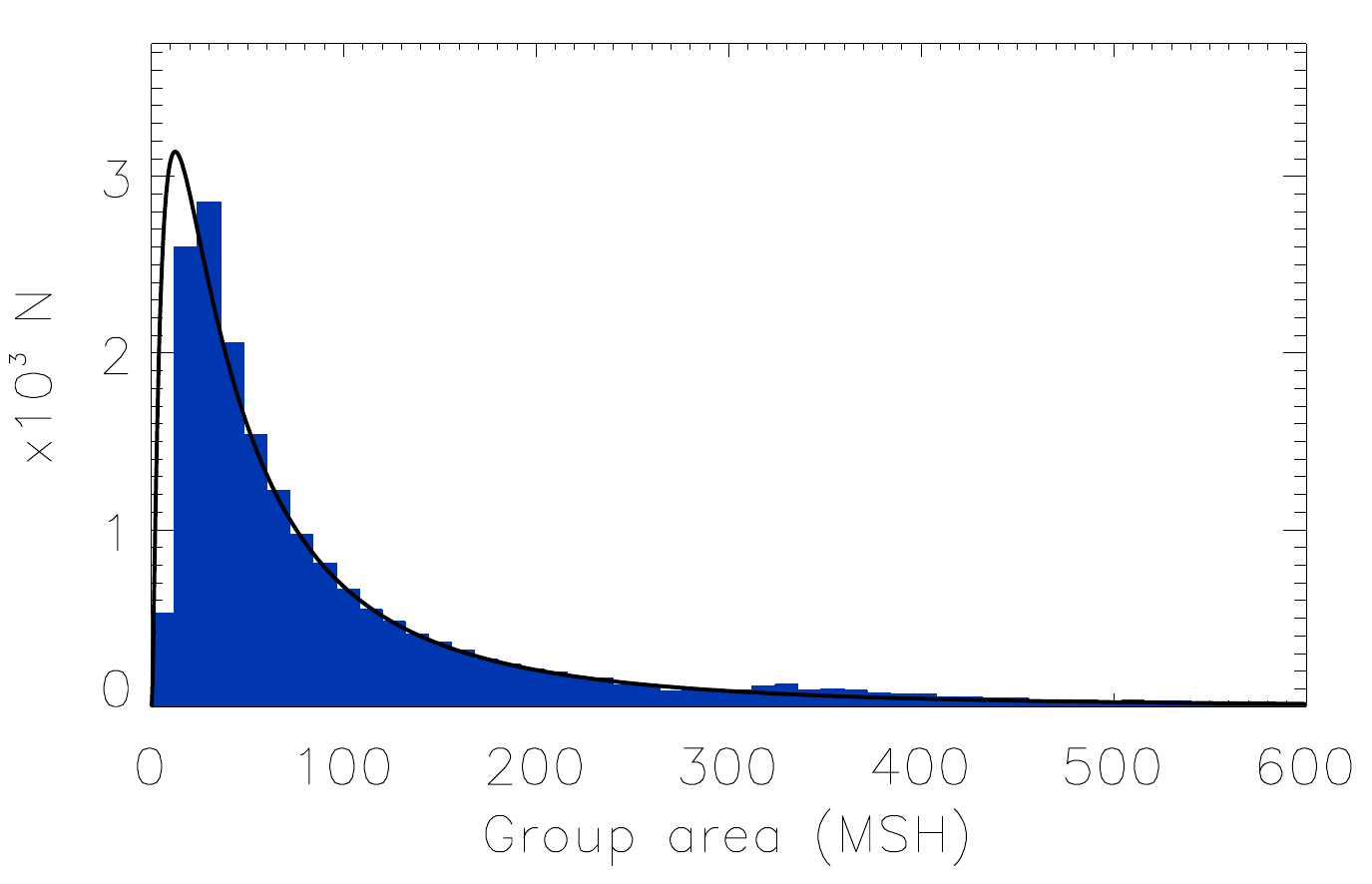}
\includegraphics[width=8.cm]{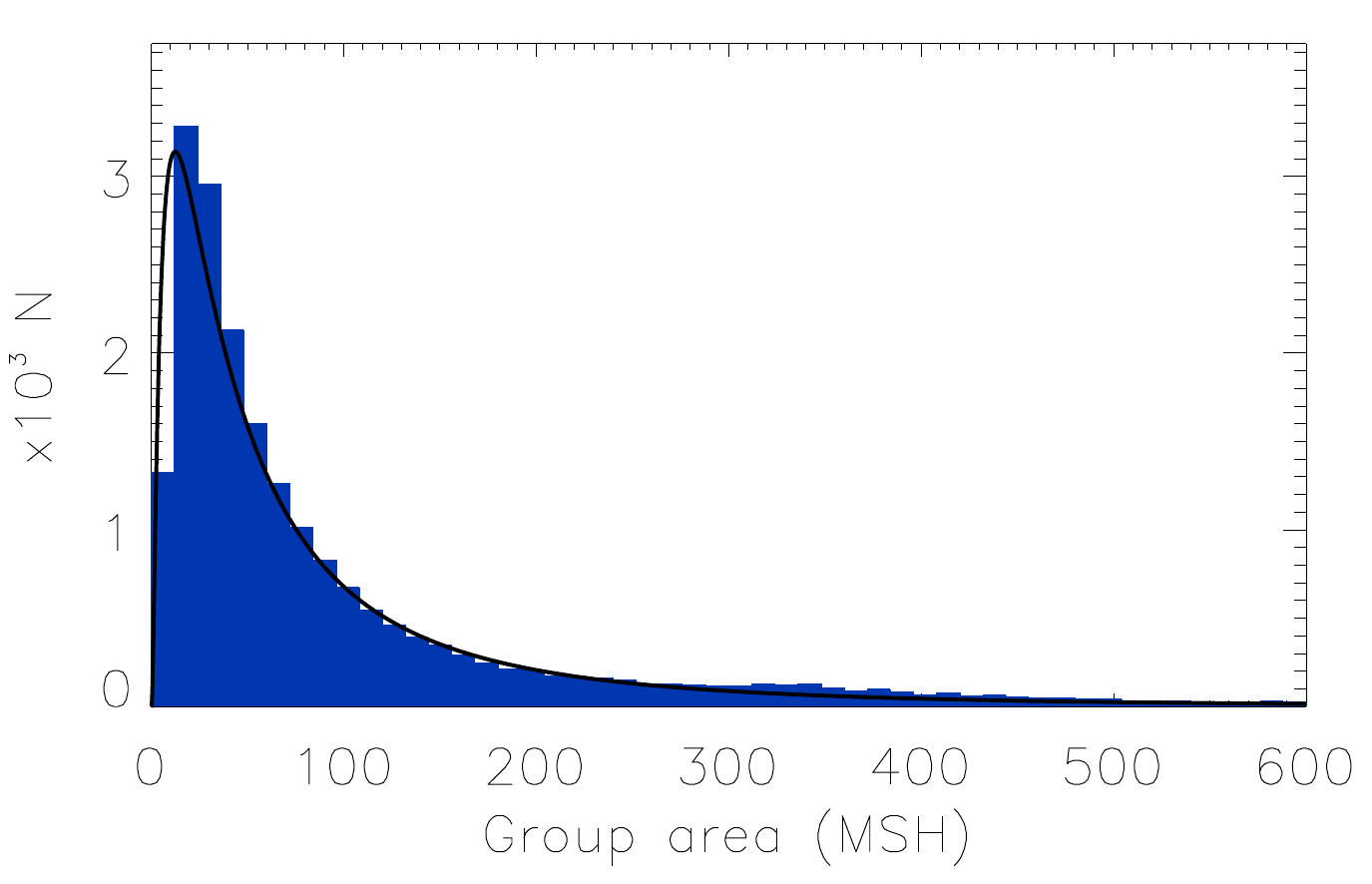}
\caption{Accumulated area distribution for synthetic sunspot groups when assuming the GW rule for all areas (upper panel) and  when using the modified area-lifetime relation (lower panel). The black, solid line is the log-normal fit to the observational data. \protect \footnotemark}\label{fig:histw}
\end{figure}

With the sunspot maximum area in hand, we can in principle determine the group's lifetime through the GW rule. However, we have found that when the GW rule is taken for all ranges of areas, the accumulated area distribution obtained with the empirical tool is not in agreement with the observed distribution: the number of small groups in the synthetic distribution is lower than that found in the Sun and the peaks of the two distributions do not coincide. The upper panel of Fig. \ref{fig:histw} illustrates this disagreement.

\footnotetext{The bin for the smallest groups is incomplete, as the sunspot records are limited to areas larger than 10 MSH.}

\begin{figure}[h!]
\centering
\includegraphics[width=7.5cm]{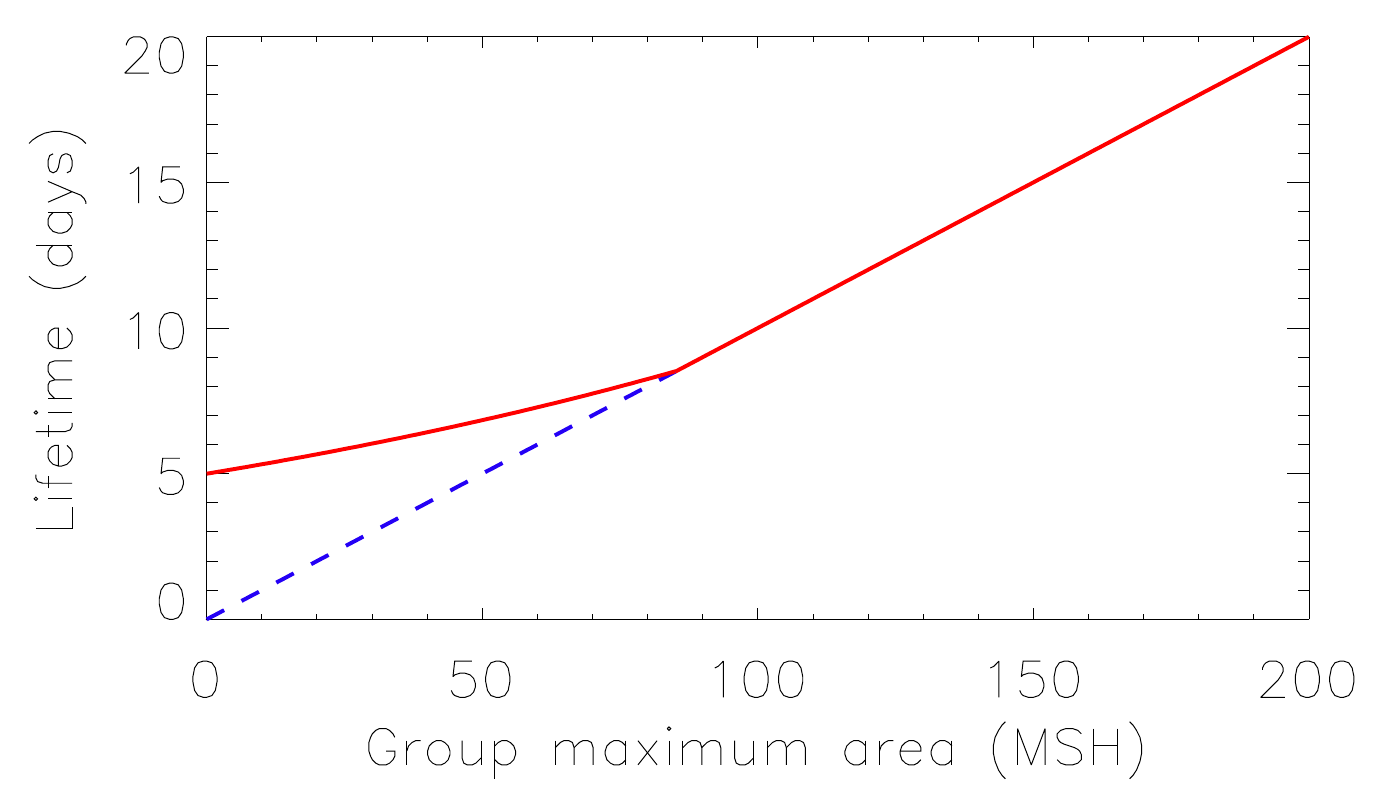}
\caption{The red line shows the area-lifetime relation assumed in the current model while the blue dashed line shows the lifetime predicted for small groups by the GW rule.}\label{fig:life}
\end{figure}

The GW rule is based on the observation of long-lived spot groups \citep[e.g. groups that live longer than $\sim20$ days;][]{henwood}. Hence, it is possible that this rule may not be adequate for the smallest groups. In fact, by tracking the small groups in the solar data for cycle 23 and comparing the time they remain visible with the lifetime predicted by the GW rule, it is possible to verify that the GW rule generally underestimates the lifetimes of the small groups. Moreover, there is a significant dispersion around the area-lifetime relation, which for the smallest groups is strongly asymmetric (since lifetimes cannot be negative). With this in mind,  we have checked whether increasing the lifetimes for the small groups would improve the agreement between the observed and synthetic accumulated area distributions and found that substituting the GW linear relation by an exponential relation at the lower areas end, our model  produces an accumulated area distribution that is in better agreement with the observed one ({lower panel of Fig.~\ref{fig:histw}}). The modification of the GW rule for groups with areas smaller than 85~MSH used in our model is
\begin{equation}
T=5\exp(6.2591\times 10^{-3}A_m).\label{eq:GWcorrect}
\end{equation}
Figure \ref{fig:life} shows the current area-lifetime relation used in our empirical model (red line).

Taking the group's maximum area and lifetime, we determine the group's area at a posterior time (while $A>0$) by applying  a decay rate of the form $\Gamma=\exp{\left(\gamma_1\right)}A^{\gamma_2}$. We then assume that the period during which the group grows corresponds to the difference between its lifetime and the decay time. In the absence of a well-established relation between the group's growth rate and its area, we opted for a power law to describe that relation, i.e. $\Psi=\exp{\left(\psi_1\right)}A^{\psi_2}$. We started by considering the values found by \citet{javariah} for $\gamma_1$ and $\gamma_2$,  making $\gamma_1=\psi_1=0.26$ and $\gamma_2=\psi_2=0.613$. These parameters were progressively changed until a reasonable agreement between the observations and the synthetic data was reached. The current version of the model considers $\Psi=\exp\left(0.17\right)A^{0.46}$ and $\Gamma=\exp\left(0.17\right)A^{0.47}$, which is consistent with the fact that large groups have higher decay rates than growth rates. For small groups the growth and decay rates given by the expressions above are essentially the same. However, since the growth time is taken as the difference between the lifetime and the decay time, in practice the smallest groups do not show a growth phase in the daily records, which might be interpreted as a fast growth, where they reach the maximum area in a time shorter than the interval between consecutive records (one day). We also tested the linear relation found by \citet{hathaway08}, but we found that decay rates are too high when compared with the lifetime of the group. Assuming the linear relation for the growth and decay rates, the time interval from the first appearance (with $A\sim0$) to the last appearance (with $A\sim0$) is much shorter than the lifetime from the modified GW rule.

\subsection{Sunspot visibility}\label{sec:rotation}

To reproduce the daily sunspot records of the Sun, we take into account the solar rotation and the fact that spots are observed only when they are on the Sun's visible side. The group's longitude is determined randomly from a uniform distribution between $0$ and $2\pi$. If this quantity is smaller than $\pi$, we consider that the sunspot group is on the visible side, otherwise we consider that it cannot be observed. Taking into account that the rotation velocity of the subphotospheric layers is not very different from that of the solar surface, we assume the following parameterization of the groups' rotation velocity, $\omega$, as a function of their latitude, $L$ \citep{snodgrass83,snodgrass90}
\begin{equation}
\omega(L)=14.71-2.33\sin^2L-1.78\sin^4L\,.
\end{equation}
Spot groups that emerge on the visible side of the Sun can move towards the invisible side and then eventually become visible again depending on their lifetimes and on the solar angular velocity at the latitude they emerge. Moreover, groups that emerge on the invisible side of the Sun can become visible. Both these facts are taken into account in our model and only groups with an area greater than $\sim$~10~MSH are considered visible (in analogy to the sunspot data).\\

A schematic summary of our empirical model is shown in Appendix \ref{sec:app1}.

\section{Results}\label{sec:results}

\subsection{Synthetic data}

The synthetic data produced with our empirical model provide information about each generated group. In analogy to what is done in the NOAA databases, the code yields the sunspot group records, which include date, latitude, group area, lifetime, position in longitude, rotation rate of the solar surface at the group's latitude, and an identification number. These records are then used to compare our results with observed data.

\begin{figure*}
\centering
\includegraphics[width=8.7cm]{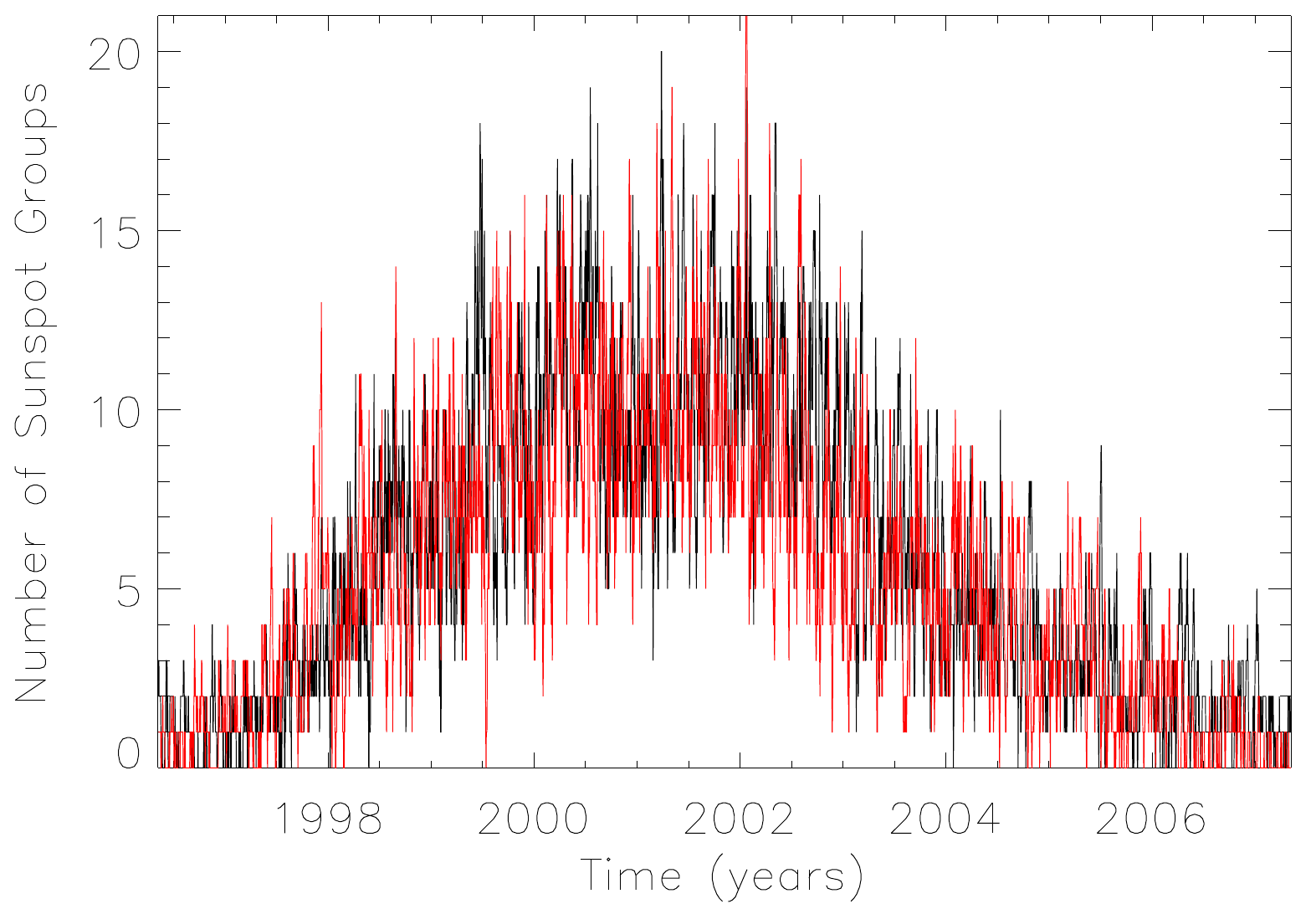}
\hspace{0.02\linewidth}
\includegraphics[width=8.7cm]{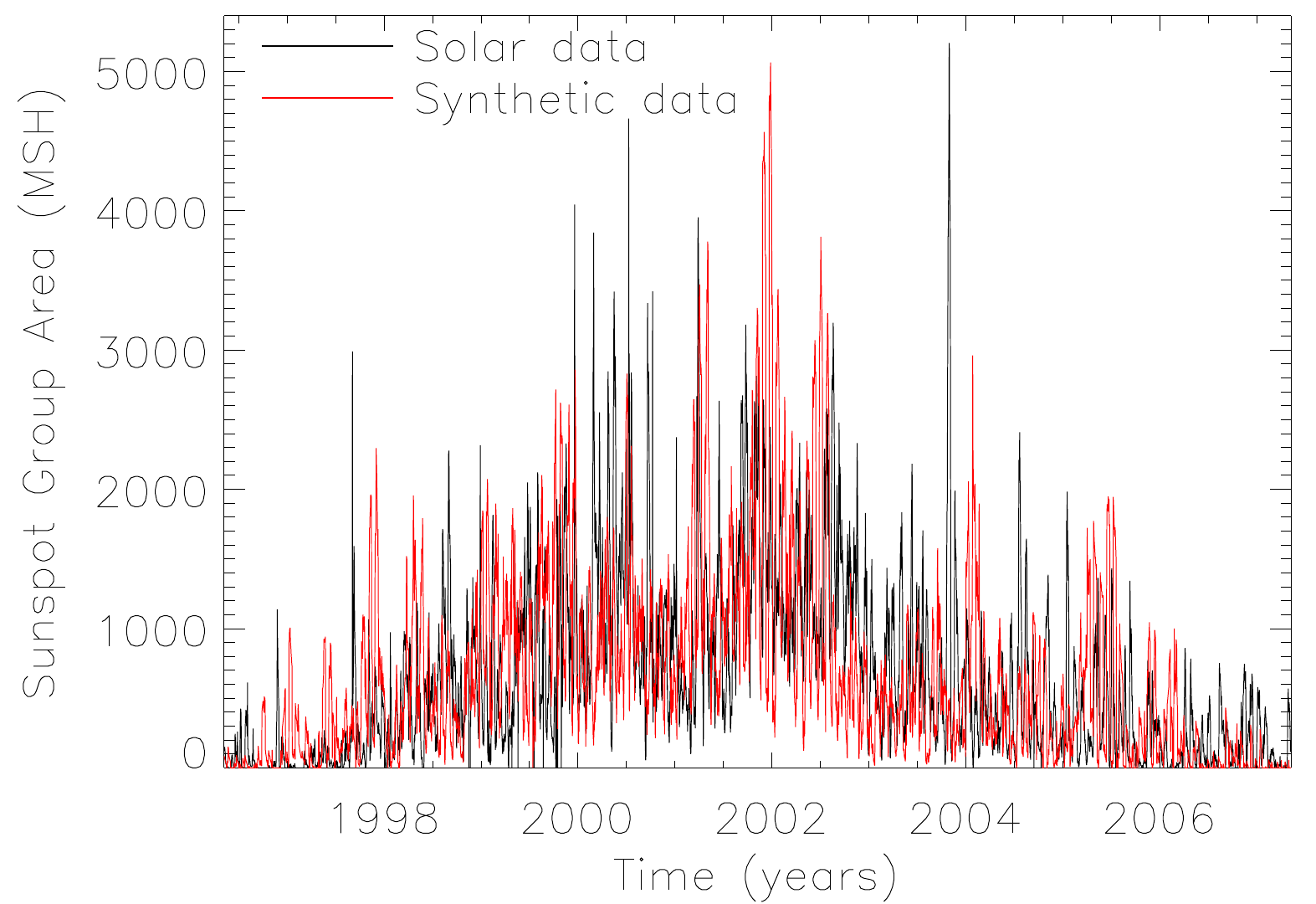}
\caption{Comparison between real solar data for cycle 23 (black) and synthetic data (red). Shown are the number of sunspot groups (left panel) and the total group area (right panel).}\label{fig:nspsimul}
\end{figure*}

\begin{figure}[h!]
\centering
\includegraphics[width=8.7cm]{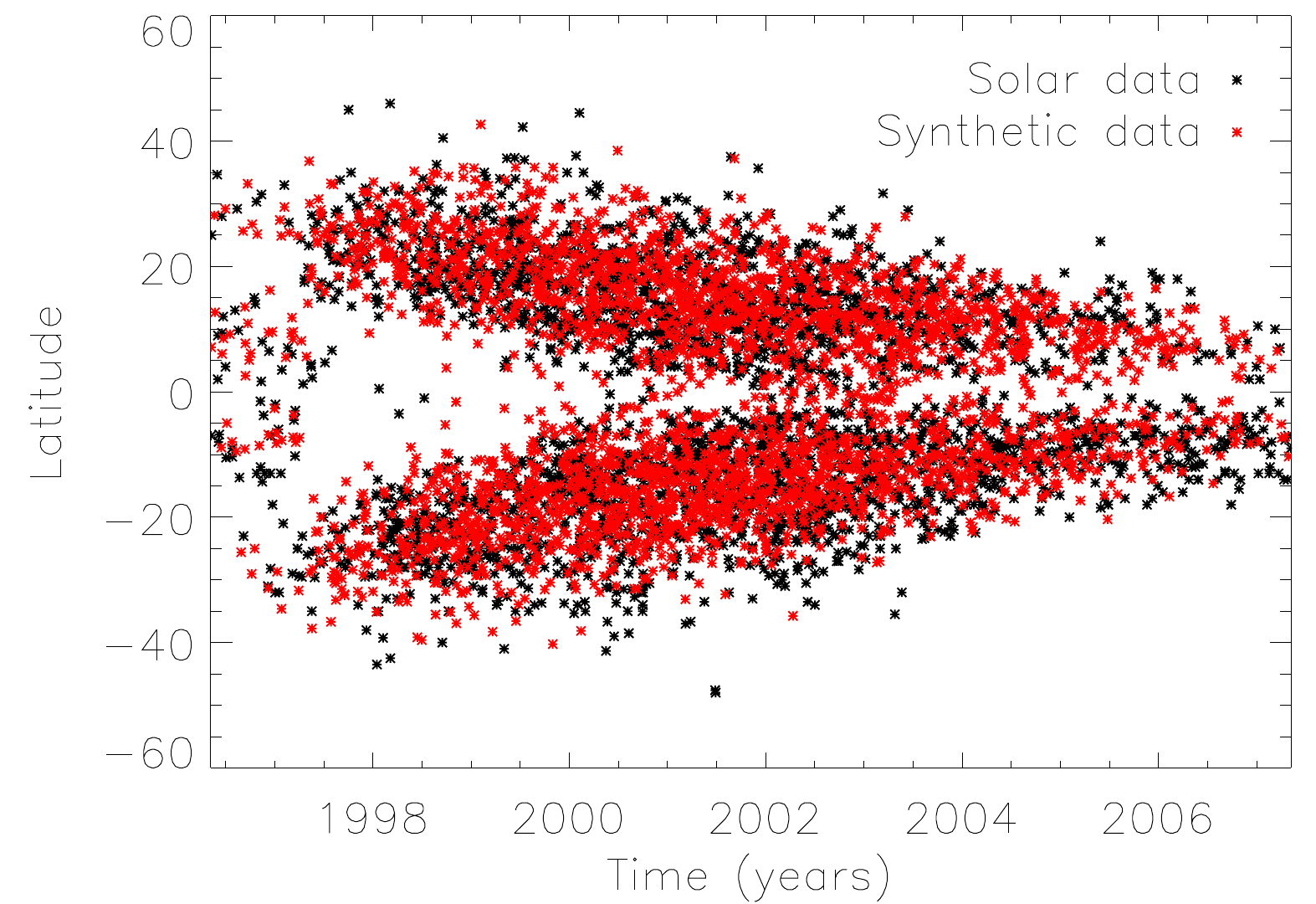}
\caption{Same as in Fig.~\ref{fig:nspsimul}, but for the real and synthetic latitudinal distribution of sunspot groups.}\label{fig:latsimul}
\end{figure}

A comparison between the number of sunspot groups observed over cycle 23 (in black) and those obtained in one realization of our model (in red) is shown in the left panel of Fig. \ref{fig:nspsimul}. The similarity in the shape and spread of the two curves is quite evident.  The same similarity is found when comparing the total group areas, i.e. the total area covered by sunspot groups in each day, (Fig. \ref{fig:nspsimul}, right panel) and the group latitudes (Fig. \ref{fig:latsimul}) for the real and synthetic data.

\subsection{Comparison test}\label{sec:KStest}

\begin{figure*}
\centering
\includegraphics[width=16.4cm]{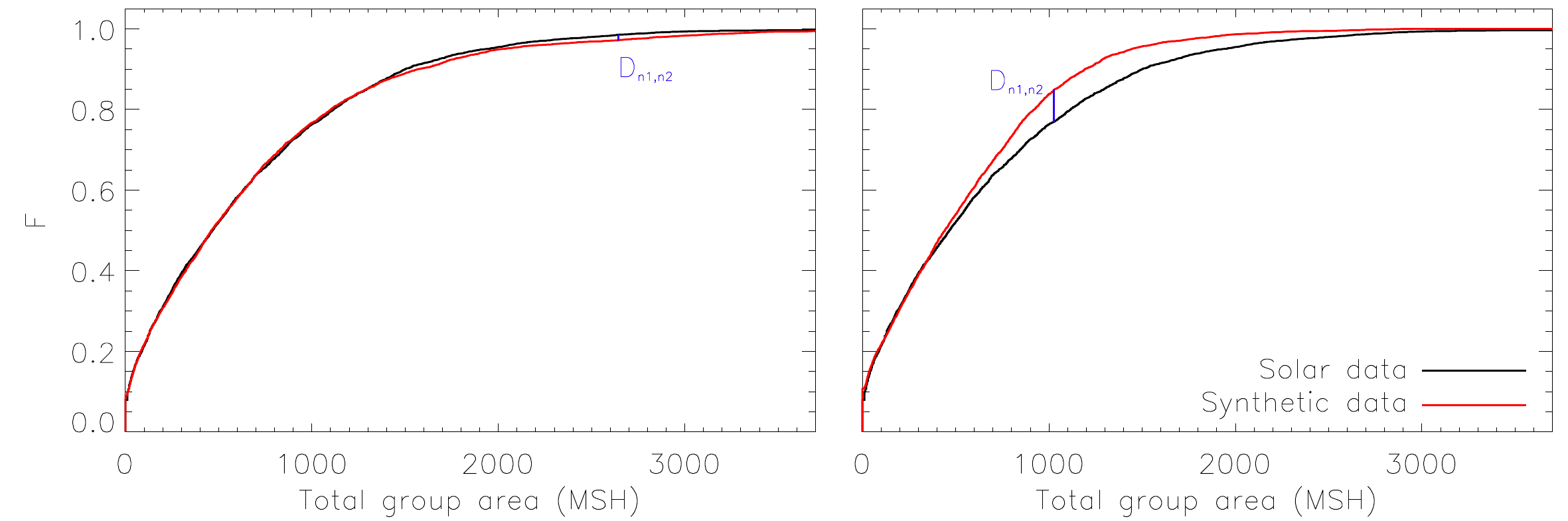}
\caption{Cumulative distribution functions of the synthetic (red) and real data (black) for the total group areas of a reconstruction obtained from the current version of the model (left panel; $D_{\textrm{n}_1,\textrm{n}_2}\sim 0.0141$) and a reconstruction obtained when assuming the GW rule for the area-lifetime relation (right panel; $D_{\textrm{n}_1,\textrm{n}_2}\sim  0.0798$).}\label{fig:kswrong}
\end{figure*}

In order to test and improve our model we have quantified how closely related the observed and synthetic data sets are by applying the Kolmogorov-Smirnov test \citep{kolmogorov,smirnov}. This test compares the cumulative distribution functions of two samples, using the maximum deviation between them,
\begin{equation}
D_{\textrm{n}_1,\textrm{n}_2}=\max_{x}|F_{1,\textrm{n}_1}(x)-F_{2,\textrm{n}_2}(x)|,
\end{equation}
where $n_1$ and $n_2$ are the number of elements of sample 1 and sample 2, respectively, and $F_1$ and $F_2$ are the corresponding cumulative distribution functions.

The null hypothesis -- i.e. that both samples result from the same distribution -- is rejected at significance level $\alpha$ if
\begin{equation}
D_{\textrm{n}_1,\textrm{n}_2}>c(\alpha)\sqrt{\dfrac{n_1+n_2}{n_1n_2}},\label{eq:kstest}
\end{equation} 
where $c(\alpha)$ is a constant that depends of the significance level to be considered.

More than to reject the null hypothesis at a given level $\alpha$, in our case this test was used to identify the aspects  of the model which needed to be improved. The synthetic sunspot cycles obtained from early versions of the model resulted in large values of $D_{\textrm{n}_1,\textrm{n}_2}$, indicating that they did not provide a good description of the observed properties of the sunspot cycle. By comparing the observed and synthetic cumulative distribution functions for the total area covered by sunspot groups and for the groups' latitudinal distribution, we could decide where and how to improve our model. An example of this is provided in Fig. \ref{fig:kswrong}, where we compare the cumulative distribution functions for the total group areas of the synthetic and real data. In this case, $F$ corresponds to the fraction of days with total area below a given value, and $n_1=n_2=4017$ is the total number of days considered in the real and synthetic data.  The left panel corresponds to a sunspot cycle reconstruction obtained with the current version of the model that considers an exponential function (equation (\ref{eq:GWcorrect})) to correct the GW rule for groups smaller than 85~MSH, while the right panel shows the results for a reconstruction obtained when adopting the GW rule for all group areas. Small groups from the former reconstruction live longer than groups with similar areas from the latter. This leads to an increase in the daily number of sunspot groups and, consequently, to a larger total area covered by the groups in the reconstruction obtained with the model that incorporates the corrected area-lifetime relation than with the other. The result is a shift of the cumulative distribution function towards larger areas and a better agreement with the observations.

\begin{figure}
\centering
\includegraphics[width=8.4cm]{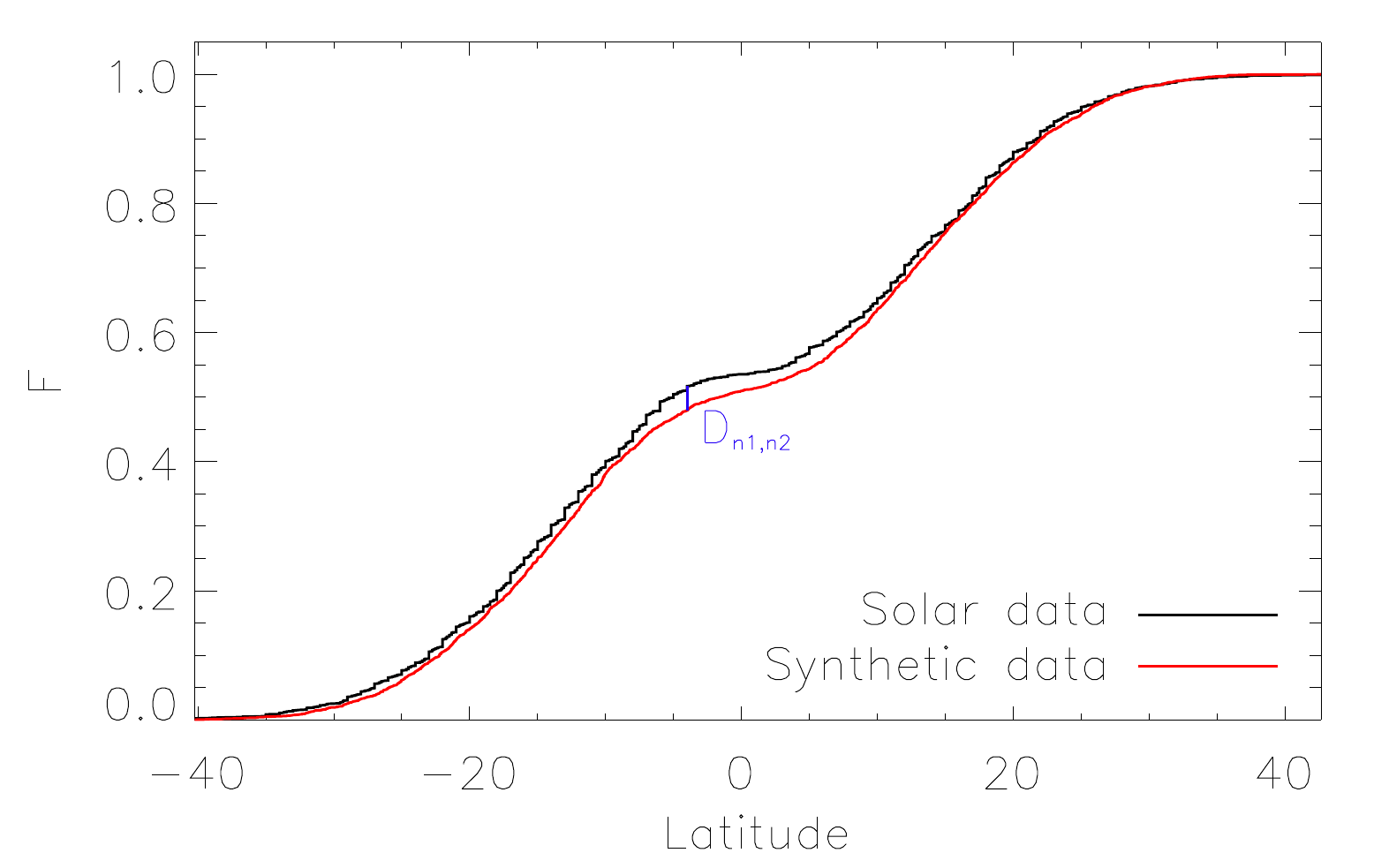}
\caption{Cumulative distribution functions of the synthetic (from the current model; red) and real data (black) for the spot latitudes. The maximum difference is $D_{\textrm{n}_1,\textrm{n}_2}\sim 0.0377$ (blue).}\label{fig:kslat}
\end{figure}

\begin{figure}
\centering
\includegraphics[width=8.4cm]{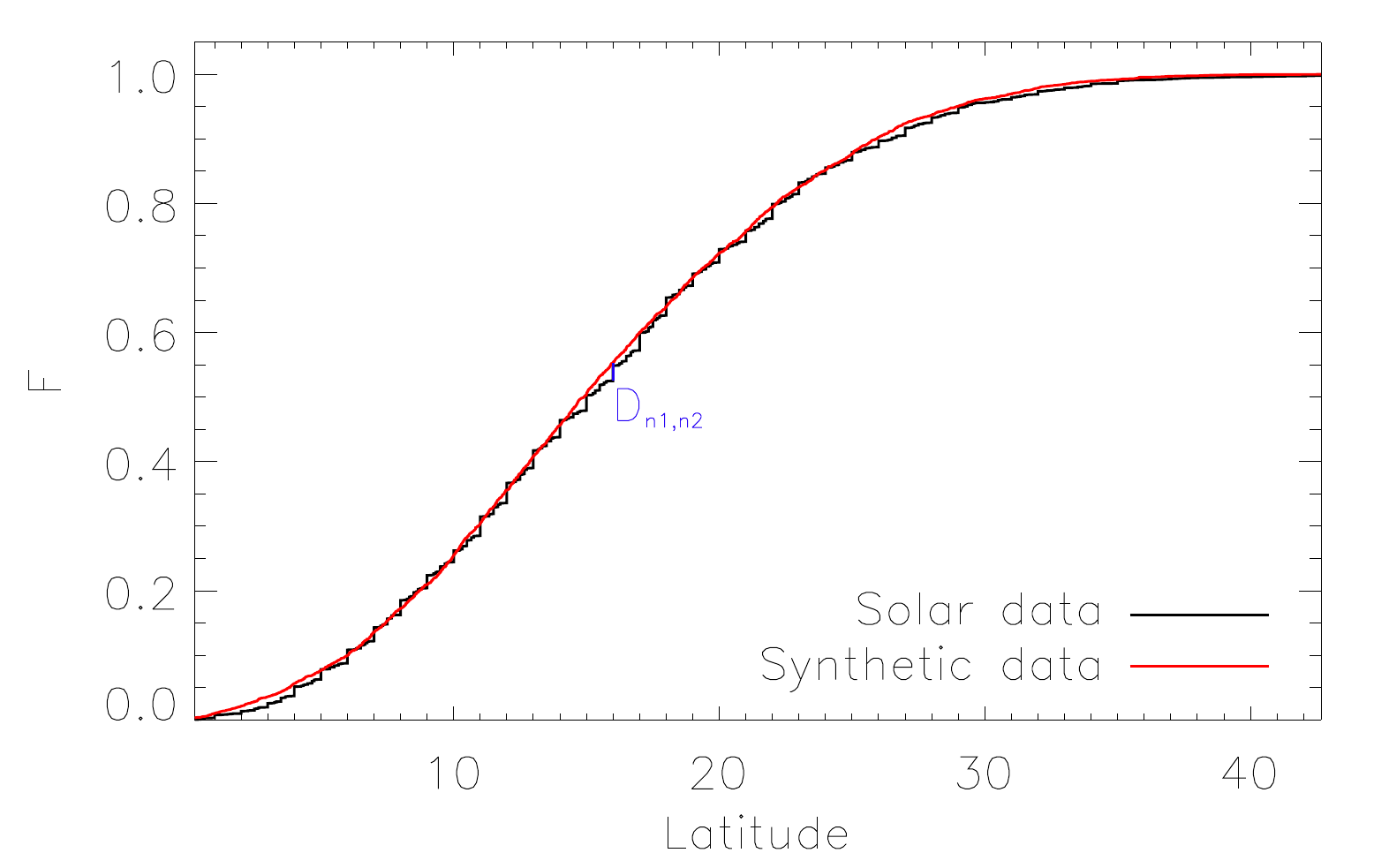}
\caption{Cumulative distribution functions of the synthetic (from the current model; red) and real data (black) for the absolute values of the group latitudes. The maximum deviation is $D_{\textrm{n}_1,\textrm{n}_2}\sim 0.0283$.}\label{fig:kslatn}
\end{figure}

Figure \ref{fig:kslat} compares the cumulative functions for the real and synthetic latitudes resulting from the same reconstruction, obtained with the current version of the empirical model. Here, the cumulative distribution functions, $F$, represent the fraction of groups that become visible (first appearance) at a latitude lower than a given value, and $n_1=2801$ and $n_2=2919$ correspond to the total number of different observed groups in the real and synthetic data, respectively. 
Despite the small value of the statistics $D_{\textrm{n}_1,\textrm{n}_2}$, the difference between the two distributions is significant and its interpretation is relatively straightforward: the cumulative distribution function for the solar data indicates that the southern hemisphere retains almost $55\%$ of the visible sunspot groups, while for the synthetic data the groups are more evenly distributed by the two hemispheres. This discrepancy results from the hemispheric asymmetry that is known to be present in the data, but that is not accounted for in our model. To verify this,  we compared the cumulative distribution functions for the absolute values of the latitude (Fig. \ref{fig:kslatn}), finding a lower value of the statistics $D_{\textrm{n}_1,\textrm{n}_2}$.

The results from the KS-test  for the synthetic sunspot cycle discussed above are summarized in Table~\ref{tb:1}. 
\begin{table*}\centering
\begin{tabular}{r|cccc}
&$n_1$&$n_2$&$D_{\textrm{n}_1,\textrm{n}_2}$&$\mathcal{D}(\alpha=0.1)$\\ \hline\hline
Total group area: Fig. \ref{fig:kswrong}, left\,&4017&4017&0.0141&0.0272\\\hline
Latitudes - Fig. \ref{fig:kslat} & \multirow{2}{*}{2801}&\multirow{2}{*}{2919}& 0.0377 &\multirow{2}{*}{0.0323}\\
Absolute Latitudes - Fig. \ref{fig:kslatn} & & &0.0283 &\\
\end{tabular}\vspace{0.3cm}
\caption{Results from the KS-test for the synthetic sunspot cycle discussed in the text. The $n_1$ and $n_2$ are the sample sizes for the real and synthetic data, respectively. $\mathcal{D}=c(${\tiny 0.1}$)\sqrt{({n_1+n_2})/{(n_1n_2)}}$ is the right-hand side of equation (\ref{eq:kstest}) for a significance $\alpha=0.1$.}\label{tb:1}
\end{table*}
At the significance level $\alpha=0.1$ the null hypothesis for the total group area and absolute group latitudes is not rejected.  Although the non-rejection of the null hypothesis does not allow us to conclude about its veracity, it certainly reinforces the expectation born from the direct inspection of Figs.~\ref{fig:nspsimul} and \ref{fig:latsimul} that the synthetic cycles obtained from our model retain the main observed properties of the sunspot cycle. 

\begin{figure}[h!]
\centering
\includegraphics[width=8.cm]{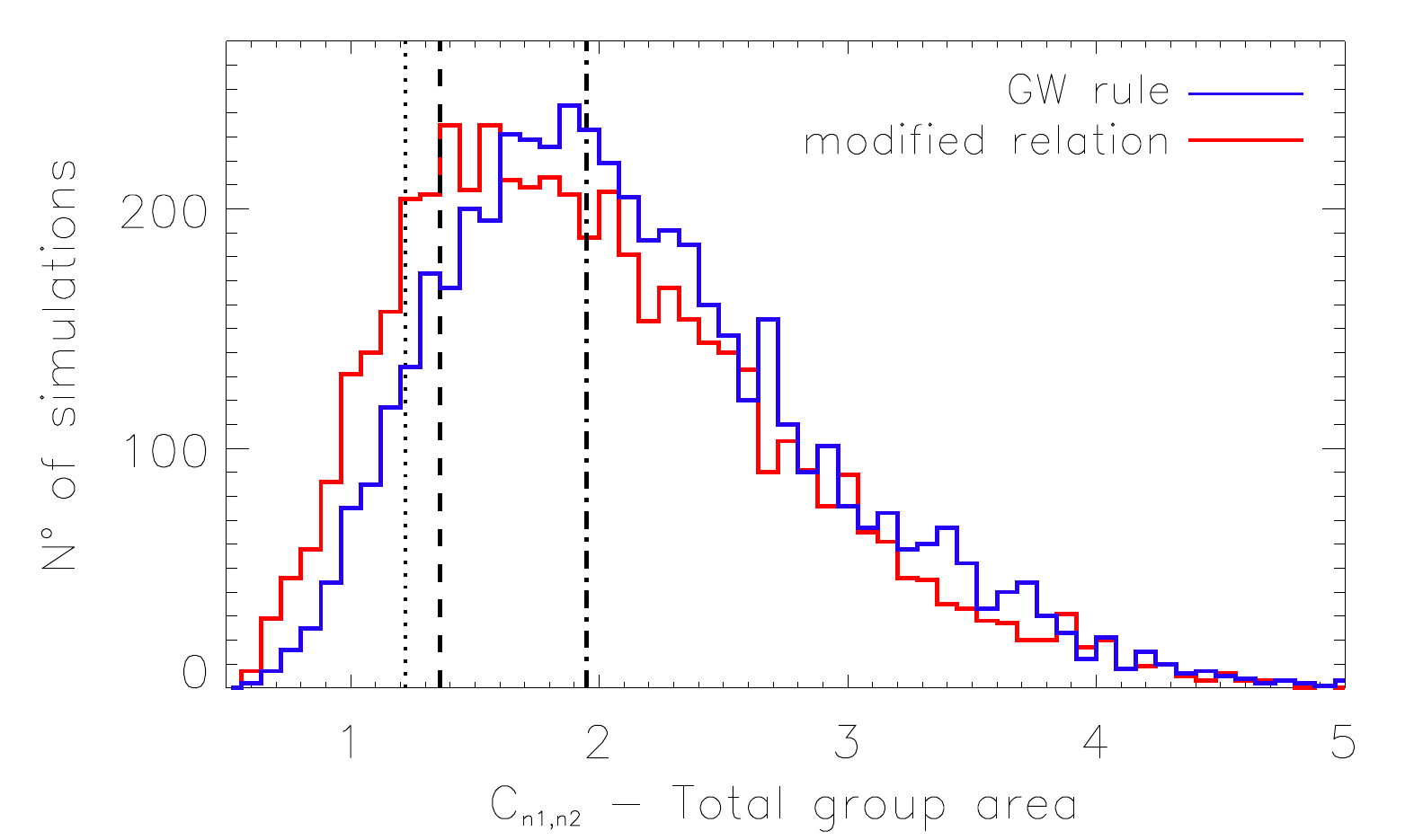}
\caption{Distribution of $C_{\textrm{n}_1,\textrm{n}_2}=D_{\textrm{n}_1,\textrm{n}_2}/\sqrt{(n_1+n_2)/(n_1n_2)}$ for the total area covered by sunspot groups, when assuming the GW rule for all areas (blue histogram) and when using the modified area-lifetime relation (red histogram). The vertical lines indicate the levels of significance $\alpha=0.1$ (dotted), $\alpha=0.05$ (dashed) and $\alpha=0.001$ (dash-dotted).}\label{fig:ddnarea}
\end{figure}

\begin{figure}[h!]
\centering
\includegraphics[width=8.cm]{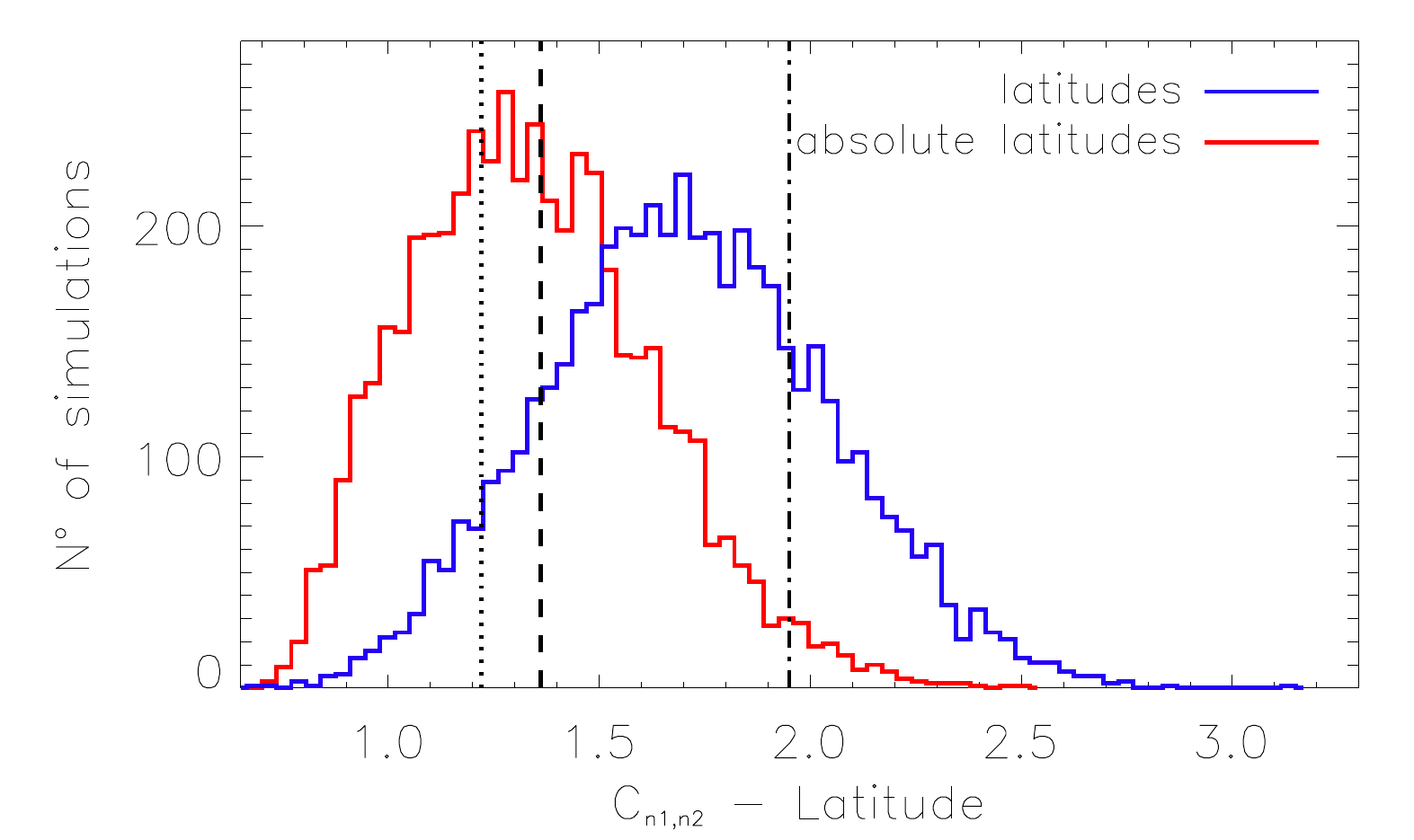}
\caption{Distribution of $C_{\textrm{n}_1,\textrm{n}_2}=D_{\textrm{n}_1,\textrm{n}_2}/\sqrt{(n_1+n_2)/(n_1n_2)}$ for the latitudinal distribution: from the group latitudes (blue) and absolute latitudes (red). The vertical lines indicate the levels of significance $\alpha=0.1$(dotted), $\alpha=0.05$ (dashed) and $\alpha=0.001$ (dash-dotted).}\label{fig:ddnlat}
\end{figure}

As the results from our empirical model are stochastic, we can perform Monte Carlo simulations to obtain the distributions for the statistics $D_{\textrm{n}_1,\textrm{n}_2}$. While $n_2$ is constant for the total area covered by sunspot groups, for the latitudinal distribution $n_2$ varies from reconstruction to reconstruction and according to the case considered (both, southern or northern hemispheres). Thus, rather than considering the distributions for $D_{\textrm{n}_1,\textrm{n}_2}$, we consider those for the $n_2$-independent quantity $C_{\textrm{n}_1,\textrm{n}_2}=D_{\textrm{n}_1,\textrm{n}_2}/\sqrt{(n_1+n_2)/(n_1n_2)}$. Figures ~\ref{fig:ddnarea} and \ref{fig:ddnlat} summarize the results obtained from 5000 cycle reconstructions. The  distribution of $C_{\textrm{n}_1,\textrm{n}_2}$ obtained from the analysis of the total group area when the GW rule is assumed for all ranges of area (Fig.~\ref{fig:ddnarea}, blue histogram) is shifted towards larger values of $C_{\textrm{n}_1,\textrm{n}_2}$ than that obtained from reconstructions that apply the correction to the area-lifetime relation for small groups (Fig.~\ref{fig:ddnarea}, red histogram). This confirms that the modified area-lifetime relation used in our model produces results that are statistically in better agreement with the solar data. Concerning the sunspot groups' latitudes (Fig.~\ref{fig:ddnlat}), the comparison of the distributions for $C_{\textrm{n}_1,\textrm{n}_2}$ clearly confirms our earlier findings. The consequence of the non-inclusion of the hemisphere asymmetry in our model is that our synthetic cycles compare significantly better when considering the absolute values of the latitude.

The properties of the sunspot emergence are cycle-dependent, stronger cycles usually having higher latitudes and wider sunspot formation zones than weaker cycles \citep[e.g.][]{solanki08,jiang}. With this in mind we have applied our model to a second cycle, namely cycle 22, which is stronger and more asymmetric than cycle 23. In agreement with the works mentioned above, we found that the average latitude and the width of the sunspot formation zone obtained in our reconstructions for cycle 22 are larger than those obtained for cycle 23. Although the results from the KS-test for the latitudes were found to be worse for cycle 22 owing to the hemispheric asymmetry, we found that with regard to the total area covered by spots and to the absolute latitudes the synthetic data for this cycle is also in qualitative agreement with the real sunspot data.

\section{Conclusions}\label{sec:conclusions}

In this work, we have presented an empirical tool for the stochastic reconstruction of sunspot cycles. With the parameters adopted in the version presented here, our tool produces synthetic daily sunspot records which retain the main properties of the real solar data.

A crucial assumption of our model is that different sunspot groups are generated independently. Despite evidence that sunspots tend to form within active longitudes \citep[e.g.][]{bumba,bogart,jiang}, pointing to possible correlations between their emergence, we found that the properties of the sunspot cycle are reasonably well reproduced under this model assumption. If a significant dependence between the generation existed we would expect that to have been reflected, for example, in a lack of agreement between the dispersion of the number of synthetic and observed sunspot groups as a function of time. In fact, we have checked that this problem would occur if individual spots were (incorrectly) considered as independent events.

An important by-product of our work was the verification that the GW rule is only appropriate for the largest sunspot groups. For the smallest sunspot groups, this rule seems to underestimate the groups' lifetime. We have proposed a modified area-lifetime relation for small groups which leads to a closer agreement of the synthetic sunspot cycle with observations.

The quantity, quality, and diversity of solar data, can only be adequately reproduced by a relatively complex model that includes a number of empirical parameters and functions describing the average observed properties of the sunspot number, area, latitude and rotation. In contrast, for other stars the observational constraints are much more limited. Thus, the application of our tool to the study of activity cycles on other stars will require the identification of the model parameters that have a significant impact on the activity-related stellar observables, including the frequency shifts (work in progress). Another potential application of this kind of tool is related to the search for exoplanets, where new strategies for reducing activity signatures in the radial velocity and transit observations can be designed.

\begin{acknowledgements}
ARGS acknowledges the support from FCT (Portugal) through the grant reference SFRH/BD/88032/2012. MSC and PPA are supported by FCT through the Investigador FCT contracts of reference IF/00894/2012 and  IF/00863/2012 and POPH/FSE (EC) by FEDER funding
through the program COMPETE. Funds for this work were provided also by the EC, under FP7, through the projects FP7-SPACE-2012-31284 and PIRSES-GA-2010-269194. TLC acknowledges the support of the UK Science and Technology Facilities Council (STFC).

\end{acknowledgements}
\bibliographystyle{aa}
\bibliography{ASantos}

\onecolumn
\begin{appendix}

\section{Schematic overview of the empirical model for the solar cycle}\label{sec:app1}
\begingroup
Figure \ref{fig:scheme} provides a schematic summary of the procedure underlying our empirical tool. The green box illustrates the detailed treatment of the evolution of a sunspot group. All the parameters included in the model have been found to be important in order to obtain results that are in reasonable agreement with the observations. In the following scheme, $\textrm{U}$, $\textrm{Pois}$, $\EuScript{N}$, and $\ln\EuScript{N}$ are, respectively, the uniform, Poisson, Gaussian, and log-normal distributions. The location in longitude of each sunspot group is represented by $O$.

\begin{figure}[h!]
\centering
\includegraphics[width=16.cm]{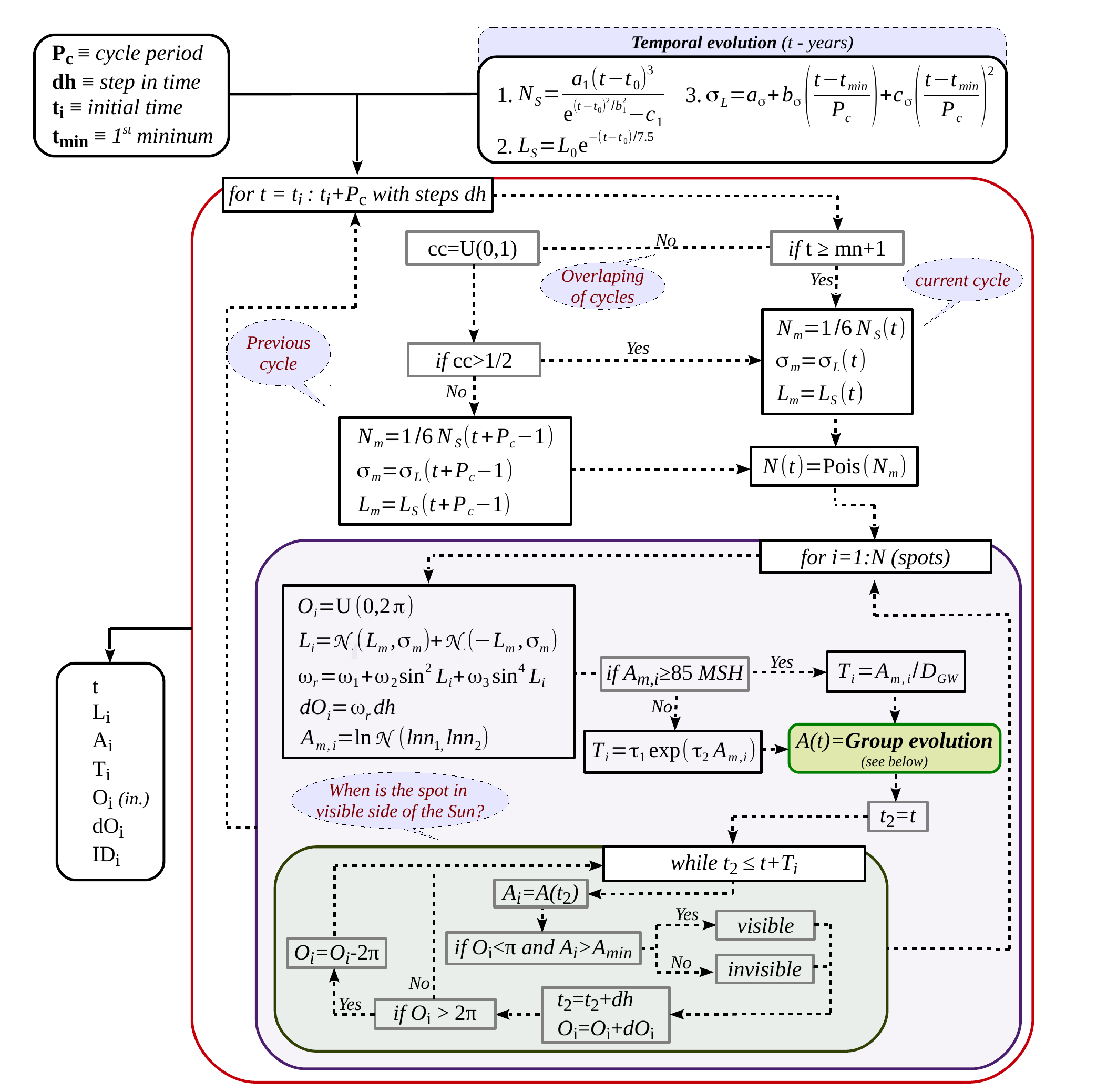}
\includegraphics[width=13.cm]{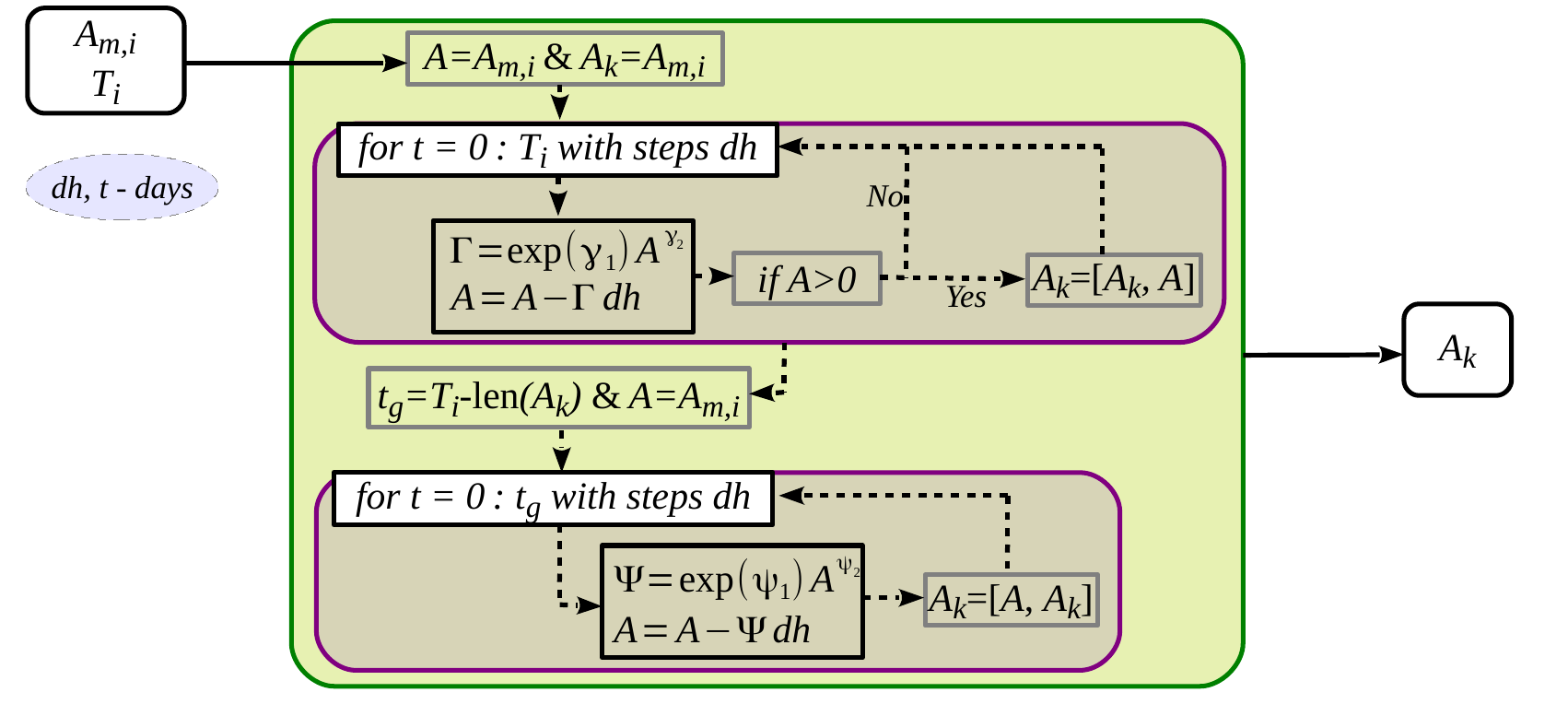}
\caption{Summary of the empirical cycle model. The green box shows how the evolution of a spot group is determined in our model.}\label{fig:scheme}
\end{figure}\endgroup
\end{appendix}

\end{document}